# Ultrafast Vortex-Core Reversal Dynamics in Ferromagnetic Nanodots

Ki-Suk Lee, Konstantin Y. Guslienko,[†] Jun-Young Lee, and Sang-Koog Kim[*]

*Research Center for Spin Dynamics & Spin-Wave Devices, Seoul National University, Seoul 151-744, Republic of Korea*

*Nanospintronics Laboratory, Department of Materials Engineering and Science, College of Engineering, Seoul National University, Seoul 151-744, Republic of Korea*

## Abstract

To verify the exact underlying mechanism of ultrafast vortex-core reversal as well as the vortex state stability we conducted numerical calculations of the dynamic evolution of magnetic vortices in Permalloy cylindrical nanodots under an oscillating in-plane magnetic field over a wide range of the frequency and amplitude. The calculated results reveal different kinds of the non-trivial dynamic responses of vortices to the driving external field. In particular, the results offer insight into the 10 ps scale underlying physics of the ulrafast vortex-core reversals driven by small amplitude (~10 Oe) in-plane fields. This work also provides fundamentals of how to manipulate effectively the dynamical switching of the vortex-core orientation.

The question of how fast magnetization (**M**) reversal occurs in bulk or finite-size magnets is one of long standing, central issues in the research field of magnetism [1,2]. In particular, the **M** dynamics of a magnetic vortex (MV) in patterned magnetic elements of submicron or less size is of growing interest because of its non-trivial static and dynamic properties [3-15] as well as promising applications to ultrafast, high-density information-storage technology [1,16]. Over the past decade the static microstructure of the MV ground state in submicron-size magnetic particles of little or zero anisotropy has been studied experimentally and theoretically [4-6], and thus is now well known [17]. The MV can be used as an information carrier, because it has two discrete states of the vortex-core (VC) orientation (up and down) and two directions of in-plane **M** rotation. From an application point of view, the switching of the VC orientation by small magnetic fields is of great importance as the potential to be used in magnetic data storage, data processing, strong spin-wave generation, and others [1,16]. However, such switching is known to be possible only with the application of a strong static magnetic field of > 2.5 kOe along the magnetization direction of VC [18]. The VC switching mode also assumes the formation and movement of a magnetic singularity (Bloch point) along the dot thickness [19]. This process demands the overcoming of a huge energy barrier, and thus is not realistic, especially at low temperatures.

Very recently experimental verification of a VC switching in Permalloy (Py) square dots driven by a small-amplitude field pulse was made by time-resolved scanning X-ray microscopy [20]. On the other hand, Hertel *et al.* gave more detailed explanation of the VC reversal dynamics by simulating the dynamic evolution of an artificial vortex-antivortex-vortex (V-AV-V) state in a Py platelet [21]. In fact, the annihilation process of the V-AV pair was firstly observed numerically in Refs. [22,23]. Through this process, considerable energy excess can be dissipated from the annihilation point in the form of spin waves. Following the idea described in Refs. [22,23], strong spin-wave radiation using a single vortex state was also suggested [24]. A similar numerical study was also carried out very recently [25].

Despite the above recent studies on the VC reversal behavior, the physical interpretation of the origin, reversal details, and VC trajectories during the reversal are still lacking. Thus, it is crucially important to explore the dynamics of the MV reversals and corresponding reversal eigenmodes in magnetic nanodots. Usually, **M** reversal is fast (~1 to 10 ns) in single-domain magnetic particles of <10 nm size and relatively slow in magnetically non-uniform particles of large size. In this Letter, we demonstrate below an important exception to this rule by calculating ultrafast **M** reversal of a non-uniform MV state in relatively large magnetic particles (of ~100 nm lateral size).

To understand the exact physical mechanism of ultrafast VC reversal under an external driving field and find a means to control it, we conducted numerical micromagnetic calculations of the MV dynamic properties in a Py cylindrical nanodot. In the micromagnetic modeling, we used a diameter of $2R = 300$ nm and a thickness of $L = 10$ nm for the Py dot. This model yield an initial **M** state of a single MV with its downward-core orientation (polarization $p = -1$) and the counter clockwise in-plane **M** rotation (chirality, $C = 1$), as shown in the top of Fig. 1. The static vortex annihilation field [9] and the gyrotropic eigenfrequency [10] for the given model are $A_s = 500$ Oe and $v_0 = 330$ MHz, respectively. Over a wide range of the field amplitude $A$ and frequency $v$, covering $A_s = 500$ Oe and $v_0 = 330$ MHz, we numerically calculated the dynamic response of the initial MV state under a time varying in-plane magnetic field $\mathbf{H}(t) = A \sin(2\pi v t)\mathbf{y}$, based on the Landau-Lifshitz-Gilbert (LLG) [26] equation of motion at $T = 0$ K.

Figure 1 shows the different dynamic regimes of the MVs in response to the driving field versus $A$ and $v$. The color-coded different regions (marked by I (gray), II (blue), III (pink), IV (yellow), and V (white)) represent a low-amplitude gyrotropic motion, the periodic creation and annihilation process of a V-AV pair, a giant-amplitude gyrotropic motion (the escape of the vortex from the dot and the vortex's return to the dot), the non-periodic process of the creation and annihilation of V-AV multi-pairs, and the instant dynamic saturation of **M** in the dot via the nucleation, movement, and annihilation of V-AV pairs, respectively. The representative trajectories of the VCs in the circular dot in the individual regions are schematically illustrated in the insets of Fig.1. Interestingly, the MV responds quite differently depending on $A$ and $v$, as shown in Fig. 1. The influence of the variable field on the MV is most effective when $v$ is close to $v_0$. For low field frequencies ($v < v_0$) and low amplitudes, the gyrotropic motion is sufficient to respond to the oscillating magnetic field. For low $v$ and high $A$, the dynamic saturation state occurs (region V in Fig. 1). In this case, even though $v$ is low, the switching process between the two saturated states occurs over a short duration (a few ns), so any kind of process can be revealed in this regime. For high $v$, the gyrotropic motion amplitude decreases as $v$ increases. Therefore, the dominant process is the deformation of the total vortex spin structure. The deformations increase with increasing $A$ and $v$, and their eigenfrequencies become higher (a few tens of GHz) than those of the gyrotropic motion. The gyrotropic motion of the VC leads to a deformation of the total dynamic spin structure of MV due to the dynamic gyrotropic field, which is proportional to the moving vortex velocity [13,27]. We observe, with increasing $A$, a gradual transition from the simple gyrotropic motion of a single MV to the multi-V-AV behavior and, eventually, to a chaotic response, through the process of the creation and annihilation of the V-AV pairs as the result of an instability of the large dynamical deformation of the vortex structure. Consequently, over a wide range of $A$ (above certain thresholds) and $v$ the ultrafast reversal of VCs occurs in general via the process of the creation and annihilation of a V-AV pair.

Here we focus on the VC reversal mediated by the periodic creation and annihilation process of a vortex-antivortex (V-AV) pair, observed in a wide range of $v$ across $v_0$ for sufficiently high $A$ values. We consider only four different conditions of ($v$, $A$) chosen due to their contrasting VC reversals, e.g., ($v/v_0$, $A/A_s$) = (4.9,1), (2.9,1), (2.4,1.2), and (2.1,1.2), as marked in the blue region of Fig. 1. First, for a case of ($v/v_0$, $A/A_s$) = (4.9,1), the snapshot images of the vortex dynamical **M** distributions and the time evolution of the different energy contributions to the total dot energy are shown in Figs. 2(a) and 2(b), respectively. The representative trajectories of the motions of VCs in the circular dot for each case of ($v$, $A$) are illustrated in Fig. 3.

Figures 2(a) and 2(b) illustrate the exact mechanism of the VC switching in relation to the dot energy and **M** distribution evolution. The V-AV pair, which has their parallel core-orientation opposite to that of the original VC orienation, nucleates, then, the original vortex ($V^{\downarrow}$) annihilates with a new AV ($\bar{V}^{\uparrow}$), and the final state is the MV of switched core orientation ($V^{\uparrow}$). From the snapshot images of the dynamic **M** distributions [Fig. 2(a)], we can clearly see each stage of the vortex reversal process. With increasing **H**, the VC velocity increases, the entire vortex structure starts to deform, the in-plane curling magnetization becomes elongated, and the VC structure (the out-of-plane **M**) becomes distorted. At $t =171$ ps, the core deformation is maximized and, then, the V-AV ($V^{\uparrow}$-$\bar{V}^{\uparrow}$) pair nucleates. Such a tri-vortex state, $V^{\downarrow}$ - $\bar{V}^{\uparrow}$ - $V^{\uparrow}$, has indeed been observed in elliptic dots in a quasi-static regime [28]. The dipolar and exchange energies (they are approximately equal during the V-AV nucleation process) reach their maximum values over the durations indicated by ③ and ④, respectively. After the moment, the nucleated AV and the original vortex move closer due to their attractive interaction [22,23], and then finally annihilate. The emission of strong radial spin waves from the annihilation site of the V-AV pair immediately follows the annihilation process [see the snapshot ⑤ in the first row of Fig. 2(a)]. Meanwhile, the newly created vortex moves away from the annihilation point. Immediately after the V-AV annihilation, the excess energy is emitted in the form of **M** waves (spin waves) [22,24]. The VC reversal is a surprisingly rapid process (sub-ns) and can be achieved in relatively low fields (~100 Oe) or extremely small fields (~10 Oe) if $v \sim v_0$. This reversal phenomenon can be understood in relation to the energy variations with field or time. In order to respond to the Zeeman torque and to decrease the Zeeman energy, the original MV structure oscillates and deforms due to the gyrotropic field. This deformation of the **M** distribution increases with increasing **H**, leading to the concentration of the dipolar and exchange energies, and hence their energy densities markedly increase in the local area near the core region (see the supplementary Figure 1 [29]). The out-of-plane **M** component in this area also increases as the in-plane **M** structure of the MV becomes more deformed. To decrease the dipolar energy, a V-AV pair eventually nucleates in this area, whereas the exchange energy decreases when the nucleated AV and the original MV structure annihilate. The deformation of the in-plane **M** structure of the MV decreases as the applied magnetic field decreases. These processes occur continuously and repeatedly in the response to the oscillating field (supplementary movies 1(a), 1(b), and 1(c) illustrate this mechanism [29]).

In contrast to the core switching occurrence described above, we also report a new finding that the VC switches or does not switch depending on the field parameters. Figures 3(a), 3(b), 3(c), and 3(d) show the individual VC reversals, with and without the core switching oppositely to the original core orientation, for time intervals equal to a half-period of the oscillating field. The process of the creation and annihilation of the V-AV pairs takes place one, two, three, and four times, respectively, as shown by the trajectories of the individual cores. The time evolutions of the VC trajectories for the cases of Figs. 3(a), 3(b), 3(c), and 3(d) are shown in supplementary movies 2(a), 2(b), 2(c) and 2(d), respectively [29]. The MV with $p = +1$ starts its motion at point ① at $H = 0$ [Fig. 3(a)], then it moves to the point ② at the moment ②. At the same moment ②, the V-AV pair with $p = -1$ nucleates at the marked site of the nucleation apart from the original MV. The newly formed AV moves to the original MV and annihilates with it, whereas the newly nucleated MV continues to move along the loop shape trajectory from the point ③ to the point ⑤. At the moment ⑤, the V-AV pair again nucleates with $p = +1$. The new AV annihilates with the old MV, whereas the new MV returns to the point ①, and this process cycle repeats in the response to the repetition of the oscillating field. Note that each of the V-AV annihilation events leads to the emission of strong radial spin waves, that is, the MV adsorbs the energy of the external field and dissipates it via the spin waves. The radial spin waves propagate to the dot border and reflect from it, forming an interference pattern [24], and then decrease their amplitude on the 100 ns time scale due to the damping in equation of **M** motion. The eigenfrequencies of the spin waves are ~ 10 GHz and determined mainly by the dot aspect ratio $L/R$ [14]. In the case of Fig. 3(b), the two-times process of the creation and annihilation of a V-AV pair is more complex than that shown in Fig. 3(a), but the principal mechanism is the same, that is, the only difference is the two-times event of the nucleation and annihilation of a V-AV pair.

Figure 3(c) shows that the reversal of the VC orientation occurs three times for the half-period and corresponds to the switch of the VC polarization, but Fig. 3(d) shows four times and no switch. If we consider that the external field has the form of a pulse field of the time duration of $1/(2\nu)$, the case of Fig. 3(a) or Fig. 3(c) corresponds to VC switching but the case of Fig. 3(b) or Fig.3(d) corresponds to no VC switching. The overall shape of the core trajectories is a zig-zag curve. The number of the turning points in each case is determined by the number of the nucleation and annihilation events of the V-AV pair during the field half-period. The VC can be switched through only the odd-numbered V-AV pair creation and annihilation events in the course of the ultrafast process of reversal of the existing single MV into a newly created one. i.e., the VC switching can be controlled by varying the field parameters and time duration.

Thus, a variable in-plane magnetic field of weak amplitude can be effectively used to manipulate the core switching of a vortex in finite-size magnetic dots. We found that this unexpected dynamic process is mediated by the creation and annihilation of the V-AV pairs in the dot. By contrast to previously studied switching mechanisms, which exploit the overcoming of some energy barrier in an applied magnetic field, we found that the switching in this case is a pure dynamic process driven by the external variable field. This allows the **M** reversal to occur on the 10 ps time scale, making it faster than any **M** reversal previously known from either experiments or calculations. The VC reversal under the influence of the oscillating in-plane field can be extremely fast (the reversal time is below 0.1 ns). The calculated VC dynamical reversal can be observed in magnetic dots of any lateral shape. The critical condition is only stability of the single MV ground state in such dots.

In conclusion, the conditions for the ultrafast reversal of vortex core and related physical mechanism involving the creation and annihilation of the vortex-antivortex pairs are identified. The new MV reversal dynamical mechanism offers a principal opportunity to use the VCs for ultrafast information recording at frequencies in the sub-THz range as well as sources of the high-amplitude spin waves. The presented calculations open a route to controlled ultrafast switching of the VCs by a small-amplitude in-plane variable magnetic field. Any variable in-plane field, a field pulse, for instance, will create the similar MV dynamic response. The ultafast VC switching is general for all dot shapes, where the static MV is stable. We believe that the VC reversal can be also reached by applying sufficiently strong spin-polarized oscillating current with frequencies close to the vortex eigenfrequency.

**Acknowledgements**

This work was supported by Creative Research Initiatives (ReC-SDSW) of MOST/KOSEF.

*Note added*: While this manuscript was being readied for submission, we became aware of a paper by Hertel *et al*.[30] which reports a similar vortex-core switching driven by a pulsed magnetic field.

# References

*Corresponding author, electronic address: sangkoog@snu.ac.kr

†Electronic address: kguslienko@snu.ac.kr

**Figure captions**

**FIG. 1** (color online). Phase diagram of the vortex dynamic response versus the frequency ($v$) and amplitude ($A$) of a time varying in-plane field in the Py cylindrical dot of $R = 150$ nm and $L = 10$ nm. The small solid spots of different colors indicate the data points in the $A$-$v$ plane where the simulations were conducted. $A_s$ and $v_0$ denote the static vortex annihilation field and eigenfrequency of the MV gyrotropic motion in the given geometry, respectively. The boundaries are drawn according to the data points where the simulations were conducted. The round- and star-shaped spots in the model dots indicate the MV and AV cores, respectively. Their blue and red colors correspond to the downward ($p = -1$) and upward ($p = +1$) core orientations, respectively. The black arrow in region V represents an instantaneously saturated **M** state.

**FIG. 2** (color online). **(a)** Underlying mechanism of the VC orientation switching mediated by the creation and annihilation of a V-AV pair and associated energy variation. The first and second rows display snapshot images of the local **M** distributions taken at the indicated times from the perspective and plane views, respectively, and the third row shows their schematic illustrations. The color and height of the surfaces in the first row indicate the local out-of-plane **M** normalized by the saturation value, $M_z/M_s$. In the second row, the streamlines with the small arrows indicate the in-plane direction of **M**. The symbols $V^{\uparrow,\downarrow}$ and $\bar{V}^{\uparrow,\downarrow}$ represent MV and AV, respectively. Their superscripts indicate the "up" (↑) and "down" (↓) core orientations. (b) Temporal variation of the exchange, dipolar, and Zeeman energies, and their sum for the core reversal of an original MV in the circular dot at $v/v_0 = 4.9$ and $A/A_s = 1.0$.

**FIG. 3** (online color). The trajectories of the individual VCs during one field period. The variation of the external field is shown in the upper row in (a) and (b). The red-, blue-, orange-, and gray-colored lines indicate the trajectories of the cores of the MV with $p = +1$, the MV with $p = -1$, the AV with $p = +1$, and the AV with $p = -1$, respectively, as marked by the symbols. The number of the creation and annihilation events of the V-AV pair is one, two, three, and four in (a), (b), (c), and (d), respectively. The small gray-colored arrows indicate the time flow and the direction of the motions of the individual vortices. In the bottom panel in (a) and (b), each open circle on the trajectories indicates the positions of each VC at the given moments, as noted by the numbers shown in the upper row. The color of each line indicates each VC trajectory. The cross-mark-in-open-circle symbols denote the nucleation sites of the V-AV pairs at the indicated moments of ② and ⑤ in (a) and ②, ③, ⑥, and ⑦ in (b). The time intervals for the MV (AV) core trajectories are $\Delta t = 5.65 - 6.25$ ns, $\Delta t = 10.00 - 11.25$ ns, $\Delta t = 7.5 - 8.75$ ns, and $\Delta t = 7.14 - 8.57$ ns in (a), (b), (c), and (d), respectively. In the insets, the snapshot images representing the corresponding core orientations at the times marked are shown for each case.

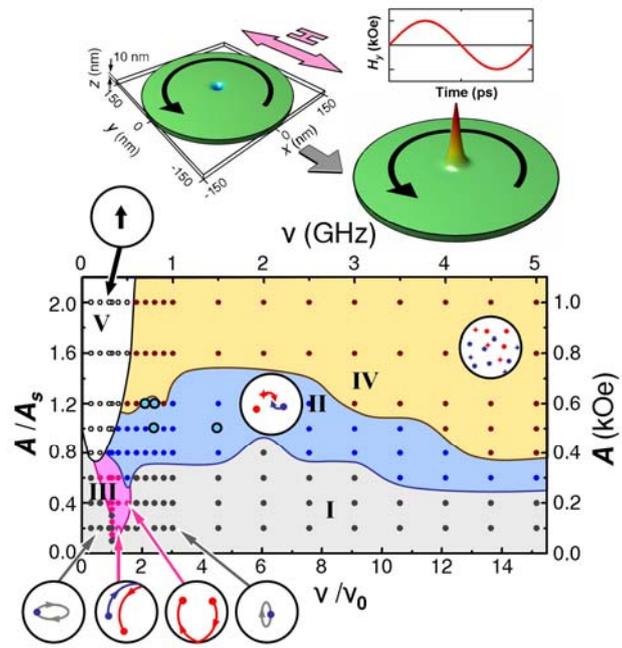

FIG.1

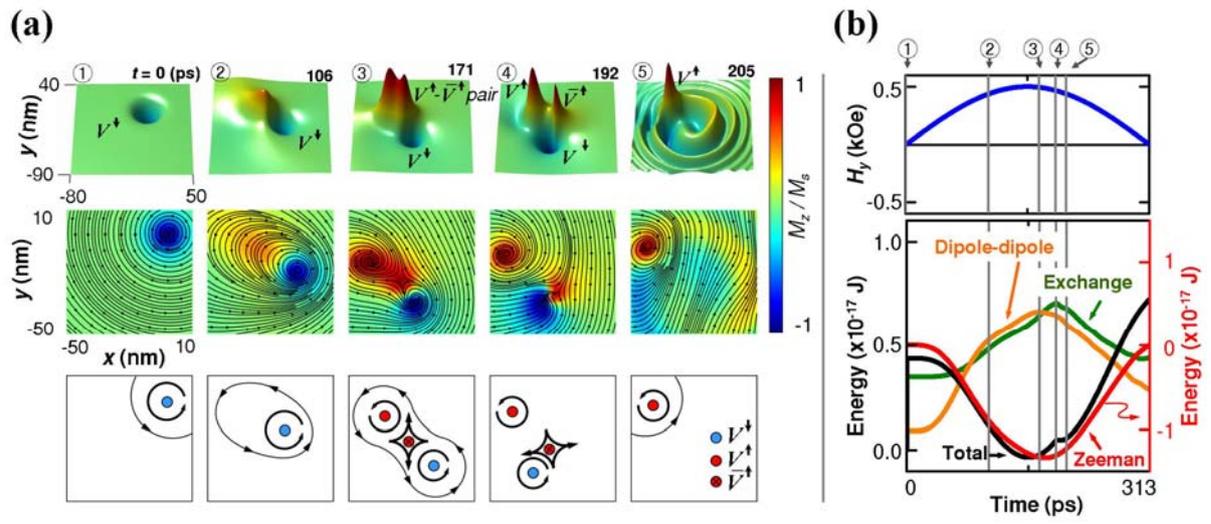

FIG.2

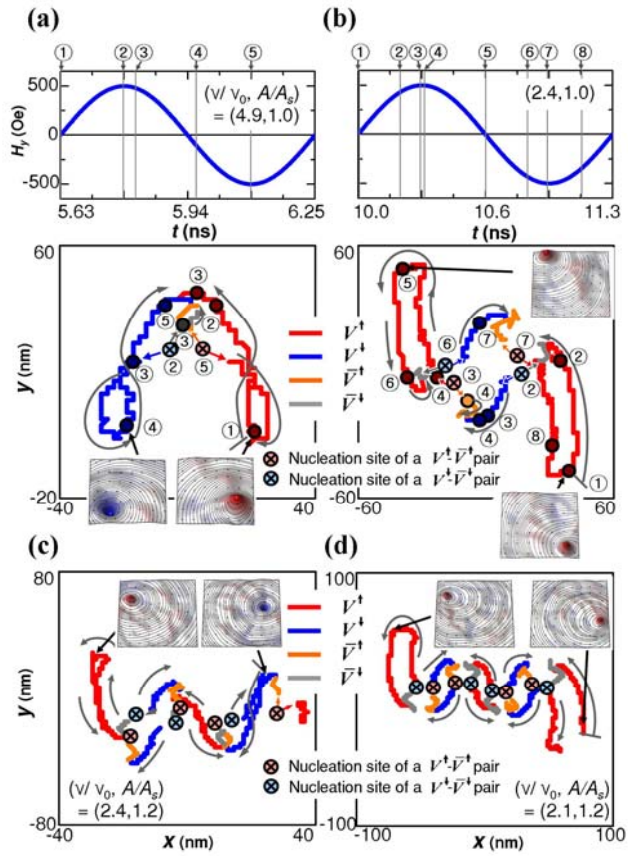

FIG.3